\documentclass[aps, pra,twocolumn,showpacs,amsmath,amssymb,superscriptaddress, longbibliography, 10pt]{revtex4-1}
\usepackage{ulem}
\usepackage{tabularx}
\usepackage{colortbl}%table add color
\usepackage{multirow}
\newcommand{\tabincell}[2]{\begin{tabular}{@{}#1@{}}#2\end{tabular}}
\usepackage{tgtermes}
\usepackage[utf8]{inputenc}
\usepackage{bm}
\usepackage{color}
\usepackage{makeidx}
\usepackage{amsfonts}
\usepackage{amssymb}
\usepackage{amsthm}
\usepackage{graphicx}
\usepackage{epstopdf}
\usepackage{subfigure}
\usepackage{amsmath,amssymb}
\usepackage[colorlinks=true,citecolor=blue,urlcolor=blue]{hyperref}
\usepackage{color}

\newcommand{\bea}{\begin{eqnarray}}
\newcommand{\ea}{\end{eqnarray}}
\newcommand{\eea}{\end{eqnarray}}

\newcommand{\sumint}[1]

\begin{document}

\title{Structure and dynamics of binary Bose-Einstein condensates with vortex phase imprinting}
\author{Jianchong Xing}
\affiliation{Shaanxi Key Laboratory for Theoretical Physics Frontiers, Institute of Modern Physics, Northwest University, Xi'an, 710127, China}
\author{Wenkai Bai}
\affiliation{Shaanxi Key Laboratory for Theoretical Physics Frontiers, Institute of Modern Physics, Northwest University, Xi'an, 710127, China}
\author{Bo Xiong}
\affiliation{School of Science, Wuhan University of Technology, Wuhan 430070, China}%\affiliation{Wuhan University of Technology, School of Science, Wuhan, 430070, China}
%\email{boxiongpd@gmail.com}
\author{Jun-Hui Zheng}
\affiliation{Shaanxi Key Laboratory for Theoretical Physics Frontiers, Institute of Modern Physics, Northwest University, Xi'an, 710127, China}
\affiliation{Peng Huanwu Center for Fundamental Theory, Xi'an 710127, China}
\author{Tao Yang}
\email{yangt@nwu.edu.cn}
\affiliation{Shaanxi Key Laboratory for Theoretical Physics Frontiers, Institute of Modern Physics, Northwest University, Xi'an, 710127, China}
\affiliation{Peng Huanwu Center for Fundamental Theory, Xi'an 710127, China}

\date{\today}

\begin{abstract}

The combination of multi-component Bose-Einstein condensates (BECs) and phase imprinting techniques provides an ideal platform for exploring nonlinear dynamics and investigating the quantum transport properties of superfluids. In this paper, we study abundant density structures and corresponding dynamics of phase-separated binary Bose-Einstein condensates with phase-imprinted single vortex or vortex dipole. By adjusting the ratio between the interspecies and intraspecies interactions, and the locations of the phase singularities, the typical density profiles such as ball-shell structures, crescent-gibbous structures, Matryoshka-like structures, sector-sector structures and sandwich-type structures appear, and the phase diagrams are obtained. The dynamics of these structures exhibit diverse properties, including the penetration of vortex dipoles, emergence of half-vortex dipoles, co-rotation of sectors, and oscillation between sectors. The pinning effects induced by a potential defect are also discussed, which is useful for controlling and manipulating individual quantum states.

\end{abstract}

\pacs{03.75.Lm, 03.75.Kk, 05.30.Jp}

\maketitle
%%%%%%%%%%%%%%%%%%%%%%%%%%%%%%%
\section{\label{sec:level1}INTRODUCTION}
Research on collective excited states of many-body systems has attracted intensive interest. Due to the unique quantum wave nature, macroscopic quantum coherence, and artificial controllability, Bose-Einstein condensates (BECs) provide an ideal platform for fully studying nonlinear topological excitations, such as vortices, solitons, knots, and skyrmions, which are difficult to be regulated in natural systems\,\cite{PhysRevLett.83.2498, PhysRevA.61.013604, PRA.87.023603, Science.287.97, LP.29.015501, CPB.30.120303, CPL.39.070304, PRE.85.036306, PRA.102.063318, PhysRevLett.86.3934, FOP.6.46}. Multi-component BECs are promising candidates for realizing steady topological systems\,\cite{FOP.17.42501,FOP.8.302}. In binary BECs, the difference of atomic properties and the competition from inter-species and intra-species interactions bring diverse phenomena, for instance, attractive-interaction-induced collapse \,\cite{PhysRevA.63.043611}, phase separation\,\cite{PhysRevLett.77.3276, PhysRevA.80.023613} , exotic vortex structure \,\cite{PhysRevA.81.033629, PhysRevLett.105.160405, PhysRevA.86.053601,PhysRevA.85.043613}, anomalous phase transition between miscible and immiscible phases\,\cite{PhysRevLett.102.070401,PhysRevLett.107.230402}, and quantum turbulence\,\cite{PhysRevLett.105.205301}. When the repulsive inter-species interaction dominates, two components of BECs become spatial separated like two immiscible fluids such as oil and water
\,\cite{PRL.81.5718}. This phase separation can also be realized by tuning the trap potential \cite{PhysRevA.85.043602,PhysRevA.87.013625,PhysRevA.94.013602} or the particle-number imbalance \cite{PhysRevA.102.023305}. The interface between the two components deforms continuously due to the effects of Rayleigh-Taylor instability under perturbation\,\cite{PhysRevA.80.063611} and Kelvin-Helmholtz instability resulting from a finite relative velocity between two components\,\cite{PhysRevB.81.094517}. Since the phase-separated BECs can be treated as a junction in atomic devices, it is also a good platform to investigate transport and tunneling dynamics of topological quasi-particles. %such as vortices\,\cite{PRA.87.023603,PRA.88.043602}. %with complex knot structures and topological states such as vortices, \textcolor{red}{skyrmions, and hofpions by tuning spin-orbit coupling\,\cite{PhysRevLett.103.250401,PhysRevLett.108.035301} and atom-atom interactions\,\cite{PhysRevLett.101.040402, RevModPhys.82.1225} that are difficult to be regulated in natural systems.

In experiments, besides mechanical rotation\,\cite{PhysRevLett.84.806,PhysRevLett.85.2223} and Raman interaction \cite{PhysRevLett.103.250401,PhysRevLett.108.035301}, topological states can be intentionally created by using phase imprinting techniques\,\cite{PhysRevLett.89.190403,SR.6.29066}. %The study of the dynamics of these topological states is an important avenue for investigating the quantum transport properties of superfluid systems.
In binary BECs, the angular momentum set by phase imprinting will be redistributed in dynamics. The associated centrifugal force may change the dynamics and phase separation of the system. For instance, during the miscible-immiscible transition of the binary BECs induced by quench of the interspecies interaction, if a vortex is initially present in only one of the species, there is no vortex in the system after reaching immiscible structure, and the dynamics of the system show different properties for the initial vortex state appearing in different species\,\cite{PhysRevA.96.043603}. In a uniform system, it was found that a vortex dipole can penetrate the interface between the two components when the incident velocity of the dipole is sufficiently large \cite{PhysRevA.85.023618}. Moreover,
vortex-antivortex superposition state may exhibit peculiar petal-like structures and asymmetric separated phases, where the vortex and anitvortex may turn into ghost vortices to break the condensate clouds \cite{PhysRevLett.95.173601,PhysRevA.77.053825,PhysRevA.87.033604}. It showed that the characteristic behaviors of counter-rotating vortices in miscible two-component Bose-Einstein condensates are completely different from those of multi-quantum vortices in single-component Bose-Einstein condensates\,\cite{PRA.88.063617}.

%The initial vortex states can be introduced into phase-separated binary condensates by phase imprinting\,\cite{PhysRevLett.89.190403,PRA.87.023603,SR.6.29066}.

In this paper, we focus on the fundamental questions that how the phase-separated structure is modified by the initial vortex phase imprinting and how the vortices evolve in dynamical processes. In Sec.\,\ref{sect2}, we describe the model for binary BECs with vortex phase imprinting. In Sec.\,\ref{sect3}, after introducing a single vortex phase imprinting to one component of the binary BEC, we study the initial density structure of binary BECs and the pinning effects induced by a trap defect which is useful for capturing and detecting the vortex cores. In Sec.\,\ref{sect4}, we introduce the phase imprinting of a vortex dipole\,\cite{PRL.104.160401}, which can be regarded as a basic topological structure carrying linear momentum. We show that binary BECs with respective vortex dipoles can display rich phase structures inducing Matryoshka-like, sector-sector and sandwich-type density profiles. Moreover, the dynamics of the systems show varying properties within different density structures. The conclusion is outlined in the last section.

\section{Model and Simulation Scheme}\label{sect2}
We consider a two-component BEC of $^{87}$Rb atoms trapped in a harmonic potential. At sufficiently low temperatures, the mean-field approximation can be applied. The order parameter field $\psi(\bm{r},t)$ satisfies the coupled Gross-Pitaevskii (GP) equations
\begin{equation}
i\hbar\partial_{t}\psi_{k}=\Bigg[-\frac{\hbar^2}{2m}\nabla^2+V_k(\bm{r},z)+\sum_{j=1,2}N_jU_{kj}|\psi_j|^2\Bigg]\psi_k,
\end{equation}
where $m$ is the atomic mass, $k=1,2$ specifies the component of the BEC, $N_j$ refers to the particle number of each component, and each $\psi_k$ is normalized to 1.
The interaction strengths $U_{kj}=4\pi a_{kj}\hbar^2/m$ can be modulated by Feshbach resonance, where $a_{kj}$ is the intra-species and inter-species $s$-wave scattering length for $k=j$ and $k\neq j$, respectively. The potential of the harmonic trap reads $V_k(\bm{r},z)=\frac{1}{2}m(\omega^2_r \bm{r}^2+\omega^2_z z^2)$, where $\bm{r}^2=x^2+y^2$ and $\omega_r$ and $\omega_z$ are the confinement frequencies in the radial and axial directions, respectively.

Throughout this article, we set $\omega_r=2\pi\times5 $ Hz and $\omega_z=2\pi\times100$ Hz. The frequency along the radial direction is much lower than the axial one so that the system can be treated as a disk-shaped quasi-2D condensate. Using the Gaussian ansatz, the wave function can be separated into
\begin{equation}
\psi_k(\bm{r},z,t)=\phi_k(x,y,t)\frac{1}{\pi^{1/4}\sqrt{a_z}}e^{-\frac{z^2}{2a_z^2}},
\end{equation}
where $a_z=\sqrt{\hbar/m\omega_z}$.
By integrating along the axial coordinate, the resulting equations for the radial wave functions in dimensionless form is
\begin{equation}\label{gpe}
i\partial_{t}\phi_k=\Bigg[-\frac{1}{2}\nabla^2+\frac{1}{2}(x^2+y^2)+\sum_{j=1,2}N_jg_{kj}|\phi_j|^2\Bigg]\phi_k,
\end{equation}
where the units of length, time and energy have been chosen to be $x_0=\sqrt{\hbar/m\omega_r}$, $t_0=1/\omega_r$, and $\epsilon_0=\hbar\omega_r$, respectively. The contact interaction between atoms becomes $g_{kj}=\sqrt{8\pi}a_{kj}/a_z$. Without loss of generality, we set the number of atoms of the two components to be the same, $N_1=N_2=1.5\times10^4$, but change the magnitude of $g_{kj}$. We focus on the phase-separated regime with $g^2_{12}>g_{11}g_{22}$ which may result in interesting phenomena.

The general wave function of the system can be written as $\phi_k(\bm{r},t)=\sqrt{\rho_k(\bm{r},t)}\exp{[i\theta_k(\bm{r},t)}]$, where $\rho_k(\bm{r},t)$ and $\theta_k(\bm{r},t)$ are the density and phase distribution for each component, respectively. In addition, it is possible to imprint a single vortex or a vortex-dipole into one component of the BEC by using the phase imprinting technique, remaining the phase of the other component unchanged. In the following calculations, the initial phase of component-2 is always set to be $\theta_2(\bm{r},0)=0$, while that of component-1 is in the form%for the single vortex case, %$\theta_1$ %is set as
\begin{equation}
\theta_1(\bm{r},0)=\prod_j{s_j\times\text{atan2}(y-y_j, x-x_j)},
\label{phase}
\end{equation}
where the function $\text{atan2}( y, x) \in (-\pi, \pi]$ %returns $\arctan( x/ y)$ excepting $ y =0$
and $s_j=\pm1$ represents the topological charge of a vortex located at $(x_j, y_j)$. We employ the split-step Crank-Nicolson Method to numerically solve the GP equation \eqref{gpe}. The initial states of the system are obtained by the imaginary time evolution.

%-----------------------------------------------------------------------------
\begin{figure}[tbp]
\begin{center}
\includegraphics[angle=0,width=0.9\columnwidth]{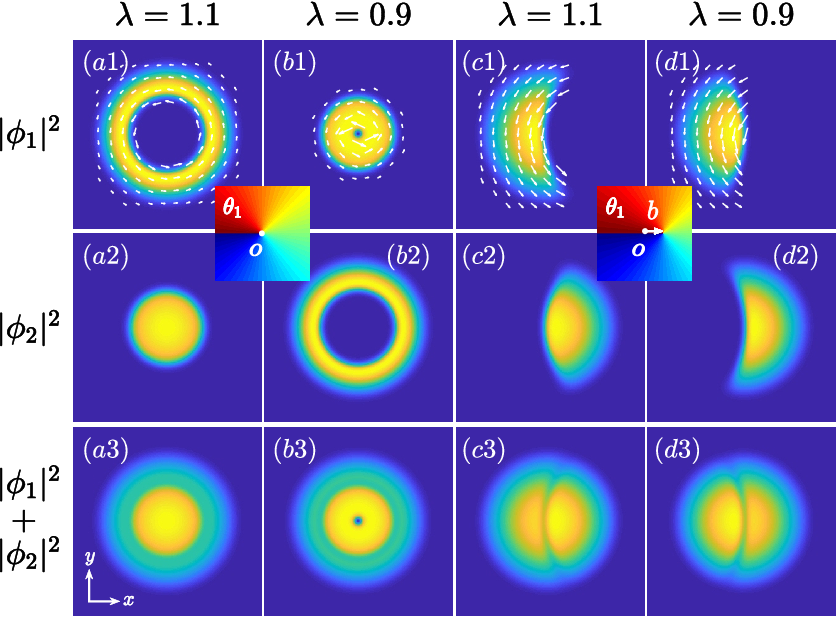}
\caption{Density profiles of binary condensates with single vortex phase imprinting in component-1. (a)-(b) The typical ball-shell structures of the density profiles with respect to different $\lambda$. The vortex phase singularity is in the trap center [$b=0$, see the inset between (a) and (b)]. (c)-(d)
Crescent-gibbous density profiles are realized when an off-center phase imprinting [$b\neq 0$, see the inset between (c) and (d)] is applied. The corresponding velocity fields for component-1 are indicated by the white arrows in the top row. Here, we set $g_{12}/g_{22}=1.5$. %The insets show the phase of component 1.
%The filed of view is $300\times300$ in dimensionless unit.
}
\label{fig1}
\end{center}
\end{figure}
%-----------------------------------------------------------------------------

\section{Single vortex state}\label{sect3}

For two-component systems in the phase-separated regime, the density distribution of each component depends strongly on the ratio of the intra-component interactions, i.e., $\lambda=g_{11}/g_{22}$, and phase imprinting applied on the system. In the situation without phase imprinting, the density distribution of the condensate shows the typical ball-shell structure \cite{PhysRevA.96.043603, PhysRevA.102.033320} as shown in Figs.\,\ref{fig1}(a1) and \ref{fig1}(a2). The total density profile shown in Fig.\,\ref{fig1}(a3) is a Matryoshka-like structure nested by the two components. The component with stronger intra-atomic interactions situated in the outer layer of the system to form a shell. For a single vortex phase imprinting, we set the distance between the phase singularity point and the trap center as $b$. For the situation of $\lambda>1$, a vortex phase {($b=0$, see the inset of the first two columns in Fig.\,\ref{fig1}) applied in component-1 will not change the structure of the condensate cloud as shown in Figs.\,\ref{fig1}(a1) and \ref{fig1}(a2). %In this part, we consider the dynamics of the system for an initial state with a single vortex imprinted in component 1 at the position $(d, 0)$. The inset of Fig.\,\ref{fig1} presents the phase distribution of component 1.
No vortex state can be identified from the density distribution. However, for $\lambda <1$, component-1 [Fig.\,\ref{fig1} (b1)] is surrounded by the shell structure of component-2 [Fig.\,\ref{fig1} (b2)], and then a vortex emerges in the ball structure of component-1 as shown in Fig.\,\ref{fig1} (b1)}. In Figs.\,\ref{fig1}(c) and \ref{fig1}(d), with applied off-center vortex phase imprinting ($b=x_0$ as an example) in component-1, the density profiles show crescent-gibbous structures in contrast to the typical ball-shell structures.  Moreover, no vortex appears in the condensate cloud no matter the value of $\lambda$ as particles in component-1 always avoid the position of the phase singularity. %\textcolor{blue}{With $d=0$???}

%In the case of $d\neq 0$, the configuration of the total density of the two-component condensate is no longer a so-called where the component with weaker interactions being surrounded by the one with stronger interactions, but a smooth arc shape as shown in .

\begin{figure}[tbp]
\begin{center}
\includegraphics[angle=0,width=0.415\textwidth]{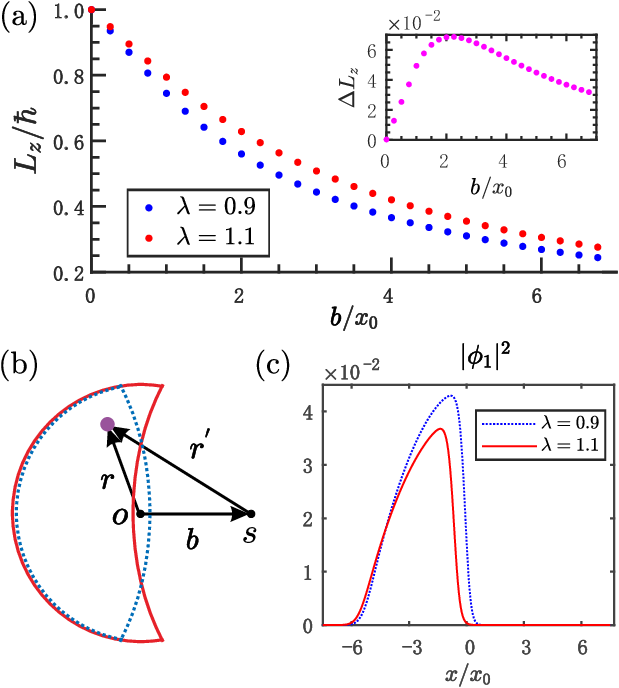}
\caption{(a) Angular momentum variation dependents on the off-center-distance of vortex for fixing $g_{12}/g_{22}=1.5$. The inset shows the value of red dot-line minus blue dot-line. (b) Density profiles of component-1 along $x$ axis. (c) Schematic illustration of component-1 of BEC to analyze the angular momentum of the system. Point $\bm o$ and $\bm s$ represent the center of the trap and vortex singularity, respectively. In (b) and (c), the red solid line and blue dashed line correspond to the parameter $\lambda=1.1$ and $0.9$, respectively.
%The filed of view is $300\times300$ in dimensionless unit.
}
\label{fig2}
\end{center}
\end{figure}

The effective action of the system is rotationally symmetric, so the angular momentum of the system,
%\begin{equation}
%\textcolor{red}{\sout{L_z =\sum_k\int\int\psi^*_k(x,y)\hat{l}_z\psi_k(x,y)dxdy\,, \nonumber}}
%\end{equation}
\begin{equation}
 L_z =\sum_k\int\int\phi^*_k(x,y)\hat{l}_z\phi_k(x,y)dxdy\,,
\end{equation}
is conserved, where $\hat{l}_z=i\hbar(x\frac{\partial}{\partial y}-y\frac{\partial}{\partial x})$. However, since the initial vortex state is not an eigenstate of the operator $\hat{l}_z$ except for the case with $b=0$, the expectation value of $\hat{l}_z$ is no longer an integral multiple of $\hbar$. As shown in Fig.\,\ref{fig2}(a), the total angular momentum decreases monotonically with increasing $b$, since the phase gradient of component-1 is smaller for a larger $b$.
{In addition, when $g_{22}$ and $g_{12}$ are fixed, the magnitude of $ L_z$ increases with $g_{11}$. We show two typical values of $\lambda=0.9$ and $1.1$ in Fig.\,\ref{fig2}(a). %which is explained in the following.
As shown in Fig.\,\ref{fig2}(b), we have
\begin{eqnarray}\label{Lz}
\bm{L} = {L}_z \hat{e}_z&=& m\int \bm{r} \times \rho(\bm r)\,\bm{v}(\bm r)  d \bm r \nonumber \\
        &=& \bm{b}\times m\int \rho\bm{v}  d \bm r  + m\int \bm{r}'\times \rho\bm{v}  d \bm r,
%v=\frac{\hbar}{mr_\perp}\kappa,
\end{eqnarray}
where $\bm{v(\bm r)}\perp \bm{r}'$ is the velocity field of component-1, and its magnitude is proportional to the inverse of the distance away from the vortex singularity point, i.e., $|\bm{v}|= s \hbar/{m |\bm{r}'|}$ \cite{note}. Therefore, the term $m \int \bm{r}'\times \rho\bm{v}  d \bm r =\hbar \hat{e}_z$ is a constant vector as the density of each component is normalized to 1, and the direction is the same as $\bm L$. On the other hand, since for $\lambda=0.9$ there are more atoms concentrating near the $x$-axis and the harmonic trap center $\bm{o}$ [see Fig.\,\ref{fig2}(b)-(c)],
the corresponding magnitude of $m \int \rho\bm{v}  d \bm r$ is larger than that for $\lambda=1.1$. As the applied velocity field in the $y$-direction is always negative [see Fig.\,\ref{fig1} (c1)-(d1)], the term $\bm{b}\times m \int \rho(\bm r)\bm{v}  d \bm r$ is antiparallel to $\hat{e}_z$. As a result, the magnitude of ${L}_z$ for $\lambda=0.9$ is smaller than that for $\lambda=1.1$ as shown in Fig.\,\ref{fig2}(a). The difference between $L_z$ for two sets of $\lambda$ is plotted in the inset of Fig.\,\ref{fig2}(a). For the limiting case with $b=0$, $\Delta L_z$ vanishes since $\bm{L} = \hbar \hat{e}_z$ for both parameter sets. For the other limiting case with $b\rightarrow\infty$, $\Delta L_z$ also vanishes as $|\bm v| \rightarrow 0$ and $\bm{L} \rightarrow 0 $ for both parameter sets.
 } %which increase in the beginning and then decreases with respect to $r$.
%-----------------------------------------------------------------------------
%----------------------------------------------

\begin{figure}[tbp]
\begin{center}
\includegraphics[angle=0,width=0.5\textwidth]{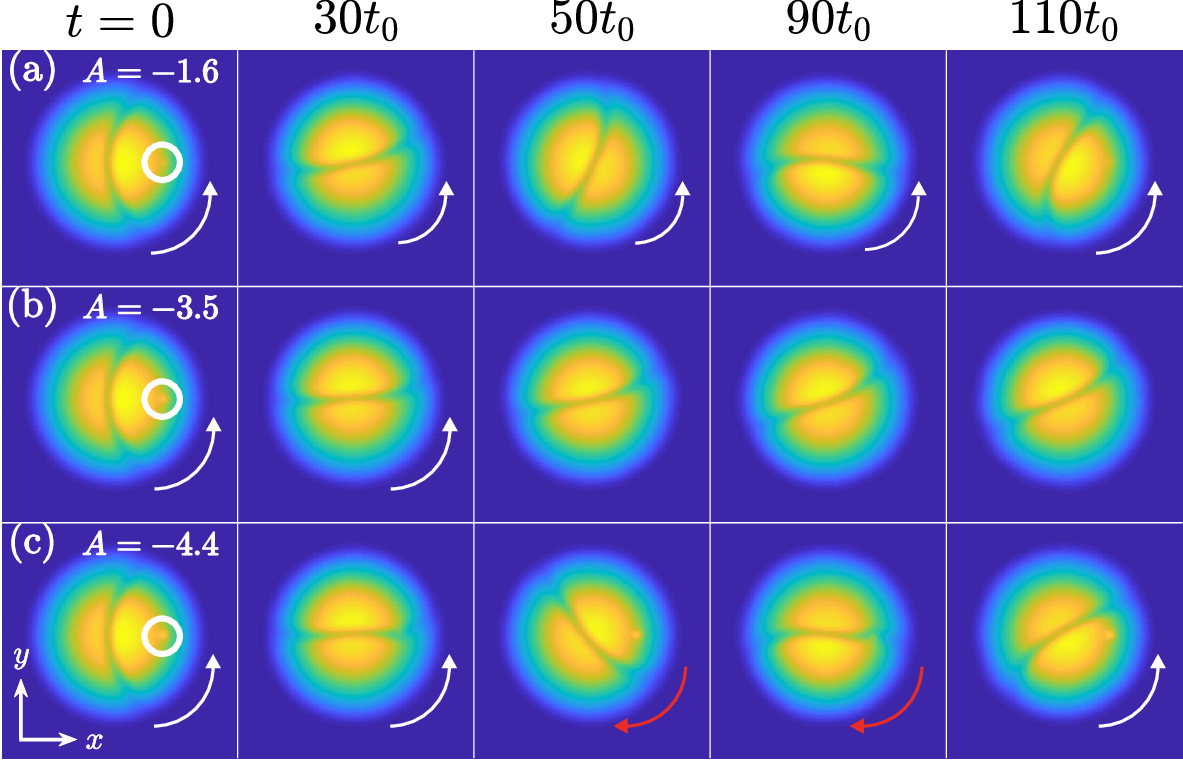}
\caption{Dynamical evolution of the density profiles after imprinting a vortex phase at position $(x_0,0)$ of component-1 and applying a weak inverted Gauss potential $V_G$ for component-2.
The white (red) curved arrow indicates that the condensate rotates counterclockwise (clockwise), while the case without the arrow indicates that the condensate is hardly rotating.
Here, the parameters used are $\lambda=1.1$ and $g_{12}/g_{22}=1.5$.
%and the red curved arrow indicates that the condensate rotates clockwise.%$(a)$ I-V corresponds the time $0, 25t_0, 50t_0, 75t_0$ and $100t_0$. $(b)(c)(d)$ I-V corresponds the time $0, 30t_0, 50t_0, 90t_0$ and $110t_0$.
}
\label{fig3}
\end{center}
\end{figure}

%--------------------------------------------------------------
In the following, we study the evolution of the total density distribution of the BEC for the initial state with a crescent-gibbous structure ($\lambda=1.1$).  We keep the form of the trap $V_1$ for component-1 unchanged, while introducing an extra Gaussian potential $V_G=Ae^{-[4(x-3)]^2-(4y)^2}$ into the harmonic trap potential $V_2$ for component-2. For $A=0$, the trap potential is unchanged for the whole system during the dynamical process, which means that the total angular momentum is conserved. Due to the existence of the initial angular momentum generated by the phase imprinting on component-1 ($s_1=1$), component-1 will push component-2 to rotate counterclockwise synchronously. %\textcolor{blue}{as shown in Fig.\,\ref{fig3}(a). The configuration of the condensates is stable and the period of the rotation is about $100 t_0$.}

For $A\neq0$, the rotational symmetry is broken by the Gaussian defect. The positions of the defect are indicated by the dashed circles in Fig.\,\ref{fig3}.
%In Fig.\,\ref{fig3}(A), we show the evolution of a smooth arc shape structure where the initial vortex located in $(x_0,0)$, i.e. $r=1$. In this case, because the vortex phase gives the component a rotational velocity field, component 1 pushes component-2 to rotate counterclockwise synchronously. The configuration of the condensates is stable and it will take time about $100t_0$ to complete a circle.
In this case, the motion of the two components %involves the mutual conversion between the kinetic energy and the potential energy,
leads to the variation of angular momentum. To investigate the pinning effects, we set $A<0$.
In Figs.\,\ref{fig3}(a)-\ref{fig3}(c), we present the motion of the BEC for different amplitude $A$ of the Gaussian potential. The initial position of the vortex is set at $(x_0,0)$. The initial values of $L_z$ for $\lambda = 0.9$ and $1.1$ are about $0.75\hbar$ and $0.8\hbar$, respectively. By adjusting the value of $A$, the condensates perform three kinds of dynamical behaviors, i.e., rotating, pinning and oscillating. For the rotating case ($A=-1.6$), the interface between the two components will pass through the defect and the two components of the BEC will continue to rotate counterclockwise. For the pinning case ($A=-3.5$), the condensate  is pinned by the defect as a metastable state after a rotation of an angle. For the oscillating case ($A=-4.4$), the condensate will be bounced back when the interface encounters the defect and then rotate back and forth.

\begin{figure}[tbp]
\begin{center}
\includegraphics[angle=0,width=0.9\columnwidth]{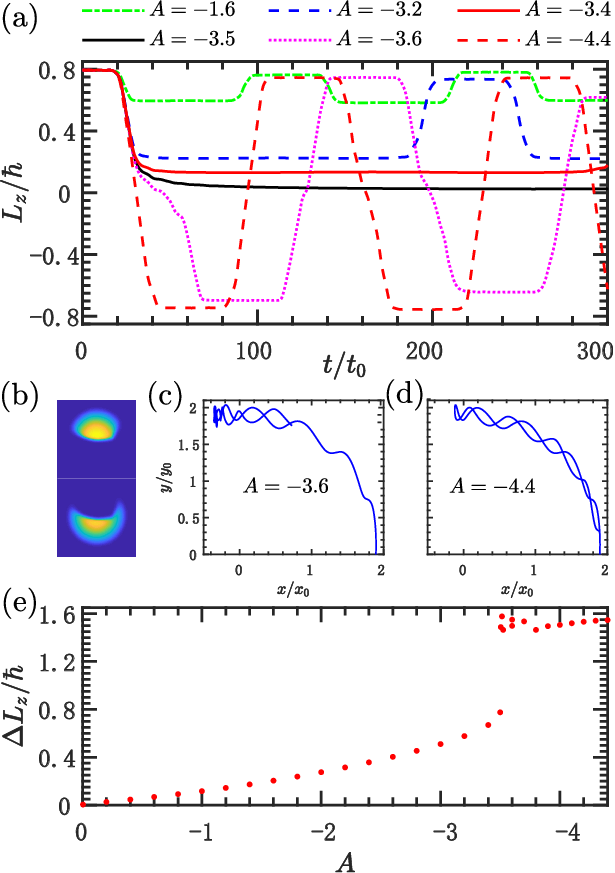}
\caption{(a) Variation of angular momentum with time after the trap adding the Gaussian potential under the parameter $\lambda=1.1$. Here, $g_{12}/g_{22}=1.5$. (b) A typical density profile of condensates when it interacts with the Gaussian defect in dynamics. (c)-(d) The center-of-mass trajectories of component-1 with the blue dot start point for $A=-3.6$ and $A=-4.4$, respectively. (e) The amplitude of the change of the total angular momentum $\Delta L_z$ within the first half period with respect to the amplitude $A$ of the Gaussian defect. %The inset shows the more detailed change of amplitude near the discontinuity.
%The filed of view is $300\times300$ in dimensionless unit.
}
\label{fig4}
\end{center}
\end{figure}

The three dynamical behaviors can be identified by the variation of the total angular momentum of the system, which is shown in Fig.\,\ref{fig4}(a). The angular momentum $L_z$ keeps unchanged when the interface is away from the Gaussian potential, resulting in the platform structures of the $L_z - t$ curves. The angular momentum starts to decrease at about $t=20t_0$ due to the fact that the interface of two components begins to passes through the defect. During this process, component-2 starts to escape the influence of the defect, which makes the increase of the potential energy (negative $A$) but decrease of the kinetic energy. As shown by the green dash-dotted line and the blue dashed line in Fig.\,\ref{fig4}(a), for the cases with a weak defect ($0<|A|<3.5$), the angular momentum will not decrease to zero, which leads to continuously counterclockwise rotation of the condensate. After component-2 is fully out of the control of the defect, the angular momentum of the system stops changing its value because component-1 is free from the influence of the defect all the time. However, When the interface passes through the defect again, component-2 enters the scope of the defect again, leading to the decrease of the potential energy but increase of the kinetic energy. Then the angular momentum will recover and the condensate turns into the next round of rotation. In this regime, the larger the absolute value of $A$ is, the larger the variational amplitude (depth of the steps) of the angular momentum is. Moreover, the period of the rotation of the condensate becomes longer.

For a strong defect ($|A|>3.5$), from $t=20t_0$, the angular momentum decreases continuously until it changes the sign, which means that the condensate is bounced back by the defect. The critical phenomenon occurs at about $|A|=|A_c|=3.5$ [see the solid red line and black line in Fig.\,\ref{fig4}(a)], the angular momentum will drop closely to zero and almost cease to change. As a result, the interface between the two components is pinned by the defect, forming a metastable state of the condensate. When $|A|$ is slightly larger than the critical value, the variation of $L_z$ is irregular at about 0 as shown by the magenta dotted line in Fig.\,\ref{fig4}(a). The time of interacting between the interface and the defect in the first period is much longer than others. We identify strong shape variation of the two components in this time interval as shown in Fig.\,\ref{fig4}(b). This can be identified by the center-of-mass oscillation of the two components. In Figs.\,\ref{fig4}(c) and Figs.\,\ref{fig4}(d), we show the center-of-mass trajectories of component-1 in the
$xy$-plane with $A=-3.6$ and $A=-4.4$, respectively. The chaotic motion of the center-of-mass during the interaction between the interface and the defect in Figs.\,\ref{fig4}(b) clearly indicates the shape variation of the condensate together with the density profiles. Moreover, for this case, the amplitude of the oscillation of $L_z$ shows obvious damping compared with other situations. In Fig.\,\ref{fig4}(e), we plot the curve of the drop of the total angular momentum $\Delta L_z$ within the first period with respect to the strength $A$ of the potential defect. There is an abrupt change at about $A= -3.5$, distinguishing the parameter range for the rotating and oscillating dynamical behaviors of the two-component BEC in the influence of the defect potential. $\Delta L_z$ increases with $|A|$ in the rotating regime. There are strong variations within $3.5<|A|<3.8$, indicating the complicated interacting process between component-2 and the defect in this region. In the regimes away from the critical point, the trend of the increase can be fitted in the form
$c_1 \exp(|A|/2)+c_2|A|+c_3$, where $c_j$ ($j=1,2,3$) is adjustable parameters. %with decreasing $A$, $\Delta L_z$ increases gradually.

\section{ Vortex dipole state }\label{sect4}

%\subsection{Vortex dipole state}
In this section, we apply a vortex dipole phase imprinting [see Fig.\,\ref{fig5}(a)] on component-1 to investigate the density structure and dynamics of the system. As the topological charges of the two vortices of the dipole are opposite, the total vorticity of the system is 0. We place the two singularity points symmetrically about $y$-axis at $(\pm d/2, 0)$. The same as in section A, two typical sets of parameters with $\lambda=0.9$ and $\lambda=1.1$ are employed. By adjusting the value of $d$ and $g_{12}/g_{22}$, abundant initial states can be obtained via imaginary time evolution. This is due to the competition of the energy between the two components. The phase imprinting of a vortex dipole will increase the energy of the condensate. In a uniform system, the energy of a vortex dipole $E_{D}$ is calculated by \cite{PS.94.075006,PhysRevA.85.023618}
\begin{equation}
  E_{D}\sim \frac{2\pi\rho\hbar^2}{m}\ln\frac{d}{x_0}\,.
\end{equation}
This means that if $\rho$ is a constant, $E_{D}$ will increase monotonically with $d$. According to the local density approximation, we can inherit the calculation of $E_{D}$ in a non-uniform system.
However, in a harmonically trapped system, the density distribution of the condensate is not homogeneous. We use the local density $\rho(\pm d/2, 0)$ of the ground state condensate to give the approximate value of $E_{D}$. As the density of the condensate decreases gradually away from the trap center with increasing gradient, the change of $E_{D}$ is no longer monotonically dependent on $d$.
For our system, $E_{D}$ shows a trend of increasing first and then decreasing, which makes the energy competition complicated and then results in abundant excited states in the regime of phase separation of the two-component system. We identify seven density profiles altogether for both kinds of $\lambda$.
\subsection{Phase diagram of the density profiles}
%--------------------------------------------------------------

\begin{figure}[tbp]
\begin{center}
\includegraphics[angle=0,width=0.45\textwidth]{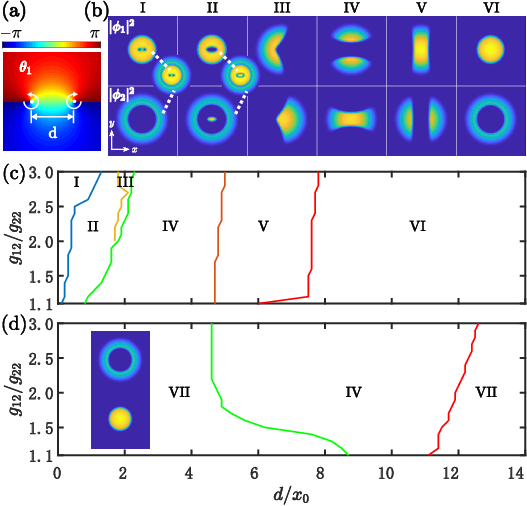}
\caption{(a) Schematic illustration of the phase imprinting in component-1 while the phase of component-2 is uniform. The singularity point of the negative vortex phase is located in $(-d/2,0)$ and the positive one is located in $(d/2,0)$. (b) Six kinds of density profiles of the condensates are realized by adjusting the value of $d$ and $g_{12}/g_{22}$ for $\lambda=0.9$ via using the vortex phase imprinting of (a). The insets in (b) indicates the total density of the two components.
The phase diagrams of the density profile of the two-component BECs with vortex dipole phase imprinting with respect to $g_{12}/g_{22}$ and $d$ for (c) $\lambda=0.9$ and (d) $\lambda=1.1$. The inset in (d) indicates the density distribution VII.
%A sandwiched density distribution is realized when the phase distribution of a vortex dipole (with vortex spacing $d$) is imprinted.
%when the phase imprinting is applied. The phase distribution of the first component of the wavefunction is determined by a vortex dipole of spacing $d$ and that of the second one keeps zero. The color lines represent the phase boundaries according to the density profile. Here, $\lambda=0.9$. %Except the border line between phase b and c contained 11 points, the rests contained 20 points each line.
%(d)The typical density profile for different phases. The first row is for the first component and the other one is for the second.
}
\label{fig5}
\end{center}
\end{figure}
%-----------------------------------------------------------------------------

In Fig.\,\ref{fig5}(b), we recognize six different density profiles labeled as I-VI for $\lambda=0.9$. We note that a vortex dipole usually lifts the energy of the condensate. Meanwhile, $g_{11}<g_{22}$ ($\lambda<1$) supports that component-2 has higher energy than component-1 if there are no vortex excitations in the system. Then component-2 would like to stay outside of component-1 to keep lower density with a shell structure, which gives a nested density profile of the system. As the phase imprinting is applied on component-1, the vortices could stay inside of component-1, which makes the energy of the two components being balanced. In this case, the density distribution of the system is given in Fig.\,\ref{fig5}(b)(I), which is also a ball-shell structure. It is intrinsically the same as the structure of one vortex phase imprinting in Fig.\,\ref{fig1}(b), but containing a vortex dipole. We can still identify the vortex dipole structure in the total density distribution of the system as shown in the inset of Fig.\,\ref{fig5}(b)(I). The core region of the vortex dipole in component-1 is not occupied by component-2.

As the size of the vortex core is proportional to the healing length $\xi$ of the condensate, it will increase when the location of the vortex is away from the center of the condensate in a harmonic trap due to $\xi\sim 1/\sqrt\rho$. Then with increasing distance $d$ between the two vortices, an oval-shaped density dip appears at the center of component-1, which makes that the vortex dipole structure cannot be identified from the density profile of component-1. Meanwhile, with increasing energy of the vortex dipole, component-2 can fill up the density dip of component-1 as shown in Fig.\,\ref{fig5}(b)(II), which is similar to the bull's eye structure formed by the one vortex quench dynamics in Ref.\,\cite{PhysRevA.96.043603}. The total density of the system shows Matryoshka-like structure as shown in the inset of Fig.\,\ref{fig5}(b)(II). For $g_{12}/g_{22}>2$, we identify a sector-sector structure of the density profile as shown in Fig.\,\ref{fig5}(b)III, which is similar to the one vortex phase imprinting as shown in Figs.\,\ref{fig1}(c) and \ref{fig1}(d).

%to minimize the total energy of the system,
%When the distance $d$ is very small, the system favors a structure as labeled by I and II in Fig.\,\ref{fig5}(b), in which the second component of the BEC forms a shell and the first one locates inside. For density profile I, the phase imprinting does excite a vortex dipole in component 1 of the BEC. This is similar to the single vortex phase imprinting as shown in Figs.\,\ref{fig1}(b1) and \ref{fig1}(b2). phase deforms to a (b) when $d$ is increased, where the density distribution of the first component has a valley in the center while the other one has an island.

By increasing $d$ further, the energy of component-1 is larger than that of component-2. The system shows a sandwich structure where component-1 is divided into two separated parts and component-2 is in the middle as shown in Fig.\,\ref{fig5}(b)IV. To keep the energy lower, the density profile of component-1 is up-down structure to avoid the locations of the phase singularities. For larger $d$, the size of the vortex dipole becomes comparable to that of the condensate, which means that the vortex dipole has less impact on the energy of component-1. This leads to smaller energy of component-1 than that of the component-2. As a result, component-1 should occupy the region with lower trap potential by higher density, which gives a transposed sandwich structure as shown in Fig.\,\ref{fig5}(b)V. To avoid the effect of the phase singularities, the density distribution of component-1 is along the $y$-direction, which makes component-2 is divided into two parts in a left-right structure. In Fig.\,\ref{fig5}(b)VI, the density profile turns to be the ball-shell structure, which means that the size of the vortex dipole   is too large to affect the energy of component-1. The parameter range of these density profiles with respect to $d$ and $g_{12}/g_{22}$ is given in Figs.\,\ref{fig5}(c).

However, for $\lambda=1.1$, only two configurations occur, i.e. IV and VII [see Fig.\,\ref{fig5}(d)VII]. The corresponding parameter range is given in Fig.\,\ref{fig5}(d). Since both phase imprinting and $g_{11}>g_{22}$ increase the energy of component-1, the phase diagram becomes much simpler than that of $\lambda=0.9$. For small and large values of $d$, the system favors a ball-shell structure with no vortex dipole appearing in the system because the vortex dipole has no effects on component-1. For a moderate $d$, the sandwich
structure IV is formed which is the same as the case of $\lambda=0.9$
due to the fact that the distribution of component-1 always tries to mitigate the influence of the phase imprinting of the vortex dipole.

\subsection{Dynamics of the metastable states}
After obtaining the possible density distributions of the excited states with vortex dipole phase imprinting, it is interesting to know the dynamics of the system. During the imaginary time evolution of the GP equation, the wavefunction stays in a subset of the many-body completed basis with fixed phase distribution determined by the phase imprinting. Therefore, the obtained excited state for a set of given parameters keeps the minimum energy only under the constraint condition (imprinted phase distribution). In the following, we investigate the temporal evolution of the excited states obtained above to see their dynamics when the constraint condition is turned off.
\begin{figure}[tbp]
\begin{center}
\includegraphics[angle=0,width=0.45\textwidth]{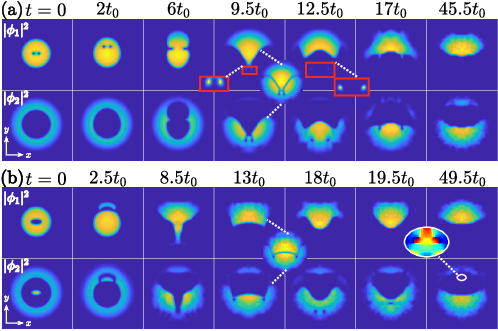}
\caption{(a) The dynamical evolution of the Matryoshka-like density profile [see Fig.\,\ref{fig5}(b)I] for $\lambda=0.9$ and $g_{12}/g_{22}=3.0$ with the initial vortex spacing being $d=1.2x_0$. The insets in the red solid oblongs magnify the corresponding regions of the density profiles.
(b) The dynamical evolution of the Matryoshka-like with component-2 being the bull's eye profile [see Fig.\,\ref{fig5}(b)II] while the initial vortex spacing is $d=1.7x_0$ for $\lambda=0.9$ and $g_{12}/g_{22}=3.0$. The insets for $t=13t_0$ and $t=49.5t_0$ are the total density of the two components and the phase distribution of the region surround by the white oval circle in the density distribution, respectively. For both (a) and (b), The first row is for component-1 and the other one is for component-2. %after the phase imprinting is withdrawn. %I-VIII corresponds the time $t=0, 2.5t_0, 13t_0, 16.5t_0, 18t_0, 19.5t_0, 38t_0$, and $ 49.5t_0$.
 %The filed of view is $6.0\times6.0$ in dimensionless unit.
}
\label{fig6}
\end{center}
\end{figure}

In Fig.\,\ref{fig6}(a), we give the dynamical evolution of the density profile I [see Fig.\,\ref{fig5}(b)I]. %The initial vortex spacing is set to be $d=1.2x_0$.
In harmonically trapped one-component condensates, the dynamics of a vortex dipole is determined by the competition between the density inhomogeneity driven and the dipole interaction driven. In two-component systems, the interaction between different species also matters. At the beginning, the center of the vortex dipole moves along the $y$ direction from the center to the periphery of component-1, and then the two vortices move apart from each other to make counter-clockwise and clockwise sense of circulations, respectively, as shown in Fig.\,\ref{fig6}(a) at $2t_0$, which is the same as that in the one-component case \cite{PLA.375.3044,SR.6.29066}. When the vortex dipole moves to the very edge of component-1, the core regions are filled by component-2 as shown in Fig.\,\ref{fig6}(a) at $6t_0$. The density distribution of component-2 varies with the shape modulation of component-1 induced by the dipole movement, but the total density profile of the two components still keeps nested structure. As the dipole moves down to the edge of component-1, the atoms of component-1 are squeezed to the upper half-plane while atoms of component-2 moves to the lower half-plane, making the total density profile of the system a sector-sector structure. Following the transition of the density profile from the nested structure to the sector-sector structure, the vortex dipole penetrates the border of the two components and enters component-2 as shown in Fig.\,\ref{fig6}(a) at 9.5$t_0$. When the vortex dipole enters component-2, its core regions are slightly occupied by component-1 [see the inset of Fig.\,\ref{fig6}(a) at 9.5$t_0$]. The combination of the vortex dipole in component-2 and the density peaks in component-1 [see the inset in Fig.\,\ref{fig6}(a)] forms half-vortex structure in the total density profile. Then the half-vortex dipole moves in the system. The vortices in component-2 disintegrate and move in the opposite directions along the circular edge as in a one-component condensate\,\cite{PRL.104.160401}, and then vanishes at the border of the two components. component-2 tries to surround component-1 again. However, due to the ghost vortex excited around component-2, component-2 cannot recover its ring-shaped density distribution as shown in Fig.\,\ref{fig6}(a) at 45.5$t_0$.

%after they then the vortices move along the periphery of the first condensate and enter in the interface, following a distortion of the interface (see $t=6t_0$). The vortex dipole will cross the interface and develop into a half-quantum vortex dipole in the second component as shown in Fig.\,\ref{fig6}(a) $t=9.5t_0$ and $t=12.5t_0$, due to the inter-species interaction. These half-quantum vortices begin to depart from each other and finally merge into the interface(see $t=17t_0$). %Meanwhile, asymmetric mosaic configuration with density wave is formed and survives for a very short time (see VII). The condensate finally goes into a sandwiched configuration and oscillates steadily (see $t=45.5t_0$)%(see VIII).

%--------------------------------------------------------------------------
%--------------------------------------------------------------------------
\begin{figure}[bp]
\begin{center}
\includegraphics[angle=0,width=0.5\textwidth]{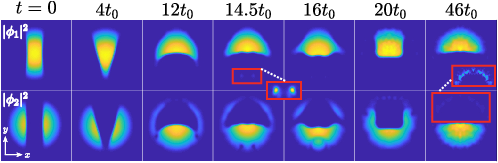}
\caption{Dynamics of the transposed sandwiched density profile [see Fig.\,\ref{fig5}(b)V] while the initial vortex spacing is $d=6.5x_0$ for $\lambda=0.9$ and $g_{12}/g_{22}=3.0$. The first row is for the  component-1 and the other one is for the component-2.
The insets in the red solid oblongs magnify the corresponding regions of the density profiles. %Here, $g_{12}/g_{22}=3.0$.
%I-VIII corresponds the time $t= 0, 4t_0, 12t_0, 14.5t_0, 15t_0, 16t_0, 20t_0$, and $46t_0$.
}
\label{fig7}
\end{center}
\end{figure}
%--------------------------------------------------------------------------
%--------------------------------------------------------------------------

In Fig.\,\ref{fig6}(b), we show the evolution of the bull's eye density profile [see Fig.\,\ref{fig5}(b)II], where the initial vortex spacing is $d=1.7x_0$. In this case, there is a density dip in component-1 and a corresponding density peak in component-2. This structure is not stable.  Even though one cannot identify the dipole structure initially, there are initial phase singularities in the phase space. The density dip is actually a composite structure of two vortices with opposite charges. During the dynamical process, the dip and peak move synchronously to the boundary of the condensate cloud as shown in Fig.\,\ref{fig6}(b) at $2.5t_0$, and disappear around the border of the two components as shown in Fig.\,\ref{fig6}(b) at $8.5t_0$. However,  the ghost vortices excited by the movement of the initial vortex dipole can enter the high density region of the condensate cloud. As shown in Fig.\,\ref{fig6}(b) at $13t_0$, a vortex dipole appears in composite 2. Unlike the process shown in Fig.\,\ref{fig6}(a), there is no counterpart of the dipole in component-1, and there is no half-quantum vortex dipole in this case. The vortex dipole moves along the periphery of the component-2 and finally combined together, leading to the formation of a soliton (see $t=19.5$). The soliton quickly merges into the interface between two components and then decays into density waves, as if it moves in a one-composite condensate. Similarly, component-2 tries to surround component-1, but its density profile is discontinuous in the upper-plane due to the ghost vortex excitations around the edge of both components [see Fig.\,\ref{fig6}(b) at $49.5t_0$]. We note that the ghost vortices can only be identified from the phase distribution of the condensates [see the inset of Fig.\,\ref{fig6}(b) at $49.5t_0$].

%In Fig.\,\ref{fig6}(b), we show the evolution of a $bull's$ $eye$ structure, where the initial vortex spacing is $d=1.7x_0$. In this case, there is a density valley in the component 1 and a density island in the component-2. The valley and island move synchronously to the boundary of the two components. A vortex dipole will appear in the component-2 as shown in Fig.\,\ref{fig6}(b) at\ $t=13t_0$. Unlike the process shown in Fig.\,\ref{fig6}(a), this vortex pair are well apart in the beginning and do not develop into a half-quantum vortex dipole. They move along the periphery of the condensate and finally combined together, leading to the formation of a soliton (see $t=19.5$). The soliton quickly merges into the interface and decays into density wave. Similarly, the condensate eventually form an oscillating sandwiched configuration ($t=49.5t_0$).
%also experiences an asymmetric mosaic configuration (VII) for a short time and eventually form an oscillating sandwiched configuration (VIII).

In Fig.\,\ref{fig7}, we show the evolution of the transposed sandwich structure with the initial vortex spacing $d=6.5x_0$. As shown in Fig.\,\ref{fig7} at $t=0$, the locations of the initial phase singularities are just outside of component-1. During the dynamics, a series of ghost vortices emerge around component-1, which makes the shape modulation of the two components spontaneously. The atoms of component-1 in the lower half-plane are pushed into the upper half-plane while the two separated parts of component-2 merge into one connected cloud. Meanwhile, component-1 is surrounded by a thin layer of component-2. There are also a vortex dipole excited in component-2 as shown in Fig.\,\ref{fig7} at $t=12t_0$. The same as in Fig.\,\ref{fig6}, the core regions of the vortices are filled up slightly by component-1 as shown in Fig.\,\ref{fig7} at $t=14.5t_0$. The dipole decays into density wave oscillations. With the continuous modulation of the interface of two component, the density profile of them turns to be about semicircles occupying the upper and lower half-planes, respectively as shown in Fig.\,\ref{fig7} at $t=46t_0$. We note that there are still slightly atom distributions of component-2 around component-1 due to the fact that the $g_{11}$ is smaller than $g_{22}$ ($\lambda = 0.9$).

%The density profile of the component 1 continuously changes its shape and it floats up gradually.  At the same time, the two parts of the component-2 of the BEC recombine together and sink to bottom as show in Fig.\,\ref{fig7} at $t=12t_0$. A half-quantum vortex-antivortex pair emerge in periphery of the  component-2. They move to each other along the periphery and induce density wave when they coincide (see $t=16t_0$). After then they move apart from each other along the periphery and finally disappear. The final condensate has a stable shape consisting of two half-moons.

\begin{figure}[tbp]
\begin{center}
\includegraphics[angle=0,width=0.5\textwidth]{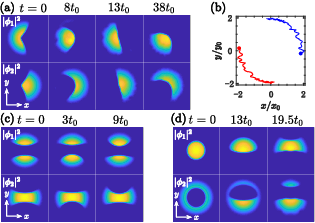}
\caption{(a) Dynamics of the sector-sector density profile [see Fig.\,\ref{fig5}(b)III] while the initial vortex spacing is $d=1.8x_0$ for $\lambda=0.9$ and $g_{12}/g_{22}=3.0$. (b) The center-of-mass trajectories of the sector-sector density profile [see (a)] where the red(blue) solid line indicates the trajectory of component-1(-2) with the start point red (blue) dot. (c) Dynamics of the sandwiched density profile [see Fig.\,\ref{fig5}(b)IV] while the initial vortex spacing is $d=2.3x_0$ for $\lambda=0.9$ and $g_{12}/g_{22}=3.0$.
(d) Dynamics of the ball-shell density profile [see Fig.\,\ref{fig5}(b)VI] while the initial vortex spacing by the phase imprinting is $d=14x_0$ for $\lambda=0.9$ and $g_{12}/g_{22}=3.0$. Here, for (a),(c) and (d), The first row is for the component-1 and the other one is for the component-2.
%after the phase imprinting is withdrawn.
}
\label{fig8}
\end{center}
\end{figure}

Other density profiles [Initial state III, IV and VI in Fig.\,\ref{fig5}(b)] are relatively stable.
For the initial condition $d=1.8x_0$, the evolution of the system starts from a sector-sector structure as shown in Fig.\,\ref{fig8}(a) at $t=0$. The interface between two components develops into a smooth arc-shape as shown in Fig.\,\ref{fig8}(a) at $t=8t_0$. Then, the density profile presents an oscillation between gibbous-crescent and crescent-gibbous like a see-saw, while the interface rotates counterclockwise at the same time. This is due to the movement of the excited ghost vortices induced by the initial phase singularities in phase space of component-1. We note that the ghost vortices emerge around component-1 first and then tunnel into the periphery of component-2. With the rotation of the system, they can move back into component-1. In Fig.\,\ref{fig8}(b), we show the trajectories of the center-of-mass of the two components by the red (component-1) and blue (component-2) solid lines, respectively. For $d=2.3x_0$, the evolution of the system starts from a sandwich-shaped structure. The density profile oscillates periodically as shown In Fig.\,\ref{fig8}(c). In Fig.\,\ref{fig8}(d), we show the evolution of a ball-shell density profile with $d=14.0x_0$. The condensate experiences an asymmetric ball-shell configuration for a short time and eventually forms an oscillating asymmetric sandwich structure as shown in $t=19.5t_0$. The number of atoms of the two parts of component-2 oscillates, and the shape of the two components varies with the modulation of the boundaries accordingly. We note that the behaviors of these two cases (initial state IV and VI) do not depend on the value of $\lambda$.

\begin{table*}[htbp]
\begin{center}
\caption{Comparison of initial structure and dynamic behavior for two-component BEC.}
\begin{tabular}{c| c| c| c| c| c}
  \hline
  % after \\: \hline or \cline{col1-col2} \cline{col3-col4} ...
  %\rowcolor{yellow}
  \tabincell{c}{Type of \\ phase-imprinting}  & \tabincell{c}{Position of the phase \\
  singularity in \\ component-1} & $\lambda$ & \tabincell{c}{Total density \\ structure} & Density profiles of two components & dynamics \\
  \hline
  \multirow{8}{*}{Single vortex}& \tabincell{c}{\\ $b=0$ \\} & $\lambda<1$ & \multirow{4}{*}{ball-shell}& \tabincell{c}{component-1: ball with vortex \\ component-2: shell} & stable \\ \cline{3-3} \cline{5-6}
  &\tabincell{c}{(centered phase \\singularity)} &$\lambda>1$& &\tabincell{c}{component-1: shell \\ component-2: ball without vortex}& stable \\
  \cline{2-6}
  &\tabincell{c}{\\ $b\neq0$ \\}
  & $\lambda<1$ & \multirow{4}{*}{Crescent-gibbous}
  &\tabincell{c}{component-1: Gibbous \\ component-2: Crescent} & \tabincell{c}{\\Rotating with interface deform \\ (amplitude of $V_G:|A|<|A_c|$)\\} \\ \cline{3-3} \cline{5-5}
  &\tabincell{c}{(off-center phase\\ singularity)} &\tabincell{c}{\\ $\lambda>1$} & &\tabincell{c}{\\component-1: Crescent \\ component-2: Gibbous} &\tabincell{c}{Pinning \\ (amplitude of $V_G:|A|=|A_c|$)} \\
  & & & & &\tabincell{c}{Oscillating \\ (amplitude of $V_G:|A|>|A_c|$)} \\ \cline{1-6}
%vortex dipole
%%%%%%%%%%%%%%%%%%%%%%11111
\multirow{20}{*}{Vortex dipole} & & &\tabincell{c}{ball-shell\\ (I)}
&\tabincell{c}{component-1: ball with vortex dipole\\ component-2: shell}
&\tabincell{c}{Transition from \\ ball-shell structure to\\ sector-sector structure with\\ formation of half-vortex\\ dipole}\\ \cline{4-6}
%%%%%%%%%%%%%%%%%%%%%%222222
& & &\tabincell{c}{Matryoshka-like\\ (II)}
&\tabincell{c}{component-1: ball with density dip\\
component-2: bull’s eye}
&\tabincell{c}{Transition from\\ Matryoshka-like structure to\\ sector-sector structure with\\ penetration of vortex dipole}\\ \cline{4-6}
%%%%%%%%%%%%%%%%%%%%%333333
&\tabincell{c}{varing $d$\\ and $g_{12}/g_{22}$} &$\lambda<1$ &\tabincell{c}{sector-sector\\ (III)}
&\tabincell{c}{component-1: sector\\
component-2: sector}
&Rotating with interface deform\\ \cline{4-6}
%%%%%%%%%%%%%%%%%%%%%444444
& & &\tabincell{c}{sandwich-type\\ (IV)}
&\tabincell{c}{component-1: two separated gibbous\\
component-2: central part}
&Oscillating periodically\\ \cline{4-6}
%%%%%%%%%%%%%%%%%%%%%55555
& & &\tabincell{c}{transposed\\ sandwich-type\\ (V)}
&\tabincell{c}{component-1: central part\\
component-2: two separated gibbous}
&\tabincell{c}{Transition from\\ sandwich-type structure to\\ sector-sector structure}\\ \cline{4-6}
%%%%%%%%%%%%%%%%%%%%%6666666
& & &\tabincell{c}{ball-shell\\ (VI)}
&\tabincell{c}{component-1: ball\\
component-2: shell}
&\tabincell{c}{Transition from ball-shell\\
Structure to oscillating\\ asymmetric sandwich\\ structure}\\ \cline{2-6}
%%%%%%%%%%%%%%%%%%%%%%
%%%%%%%%%%%%%%%%%%%%%%7777777
&\tabincell{c}{varing $d$\\ and $g_{12}/g_{22}$} &$\lambda>1$&\tabincell{c}{ball-shell\\ (VII)}
&\tabincell{c}{component-1: shell\\
component-2: ball}
&\tabincell{c}{Transition from ball-shell\\ structure to oscillating\\ asymmetric sandwich\\ structure}\\ \cline{4-6}
%%%%%%%%%%%%%%%%%%%%%444444
& & &\tabincell{c}{sandwich-type\\ (IV)}
&\tabincell{c}{component-1: two separated gibbous\\
component-2: central part}
&Oscillating periodically\\ \cline{1-6}
\end{tabular}\label{tab1}
\end{center}
\end{table*}
%III and IV.
%The behavior also does not depend on whether $g_{11}<g_{22}$.

\section{ conclusions }\label{sect5}
The combination of phase imprinting and phase separation in binary BECs results in diverse density structures and dynamics as classified in Tab.\,\ref{tab1}. The phase separation appears to minimize the system's energy when the strengths of intra-species and interspecies interactions satisfy $g_{11}g_{22}-g^2_{12}<0$. We note that this criterion is valid only at zero-temperature and in the absence of any topological defects. The phase imprinting introduced into the system adds new constrain to the competition of the energies. It opens a novel opportunity to artificially generate quantum states with vorticity in condensates and run quantum simulation of their dynamics. By adjusting the ratio $\lambda=g_{11}/g_{22}$ between the intra-atomic interactions of the two components, the ratio $g_{12}/g_{22}$ between the inter-atomic and intra-atomic interactions, and the position of the phase singularity, one can manipulate the structure of the many-body quantum states. Moreover, with the introduced potential defect, the effect of the singly-charged off-center phase imprinting can be cancelled. This can be achieved by adjusting the amplitude of the potential. The results show that the energy competition plays a key role in determining the structure of the two-component BECs.

%\break
For most of the initial density structures, to minimize the total energy of the system, the atoms tried to avoid the phase singularities, which results in no visible vortices in the condensate clouds. Vortex structures can only be identified in the ball-shell structures. The phase imprinting affects the dynamics of the binary BECs greatly. Even though the initial density profiles look the same, they will undergo different dynamical processes, such as the ball-shell structures without phase imprinting and that with vortex dipole phase imprinting while the phase singularities sit outside the condensate cloud. Most of the phase-imprinted states are not stable. The evolution of the sector-sector structure is the combination of rotation and oscillation of the density profiles of the two components, while that of the sandwich-type structure performs oscillating behaviors. All the rest unstable initial density profiles will undergo strong modulations. During the dynamics, the vortex dipole excited in the system may transfer from one component to the other. For some cases, there is transition from a vortex dipole to a half-vortex dipole.

\section*{Acknowledgments}
%This work is supported by the National Natural Science Foundation of China under grants Nos. 12175180, 11934015, 12247103 and 12247186, and the Major Basic Research Program of Natural Science of Shaanxi Province under grants Nos. 2017KCT-12 and 2017ZDJC-32, and the Scientific Research Program Funded by Education Department of Shaanxi Provincial Government under grant No. 22JK0581, and the Natural Science Basic Research Program of Shaanxi under grand No. 2023-JC-QN-0054. This research is also supported by The Double First-class University Construction Project of Northwest University.
This work is supported by the National Natural Science Foundation of China under grants Nos. 12175180, 11934015, 12247103, and 12247186, the Major Basic Research Program of Natural Science of Shaanxi Province under grants Nos. 2017KCT-12 and 2017ZDJC-32, the Scientific Research Program Funded by Education Department of Shaanxi Provincial Government under grant No. 22JK0581, the Natural Science Basic Research Program of Shaanxi under grand No. 2023-JC-QN-0054, and Shaanxi Fundamental Science Research Project for Mathematics and Physics under grant Nos. 22JSZ005 and 22JSQ041. This research is also supported by The Double First-class University Construction Project of Northwest University.

%\bibliography{E:\textbackslash\textbackslash exp1\textbackslash\textbackslash paper\textbackslash\textbackslash manuscript \textbackslash\textbackslash Refs.bib}

%merlin.mbs apsrev4-1.bst 2010-07-25 4.21a (PWD, AO, DPC) hacked
%Control: key (0)
%Control: author (0) dotless jnrlst
%Control: editor formatted (1) identically to author
%Control: production of article title (0) allowed
%Control: page (1) range
%Control: year (0) verbatim
%Control: production of eprint (0) enabled
%

% make some big mistakes in use Refs.bib. Actually it is
%right to use Refs rather than Refs.bib

\end{document}